\documentclass[preprint]{aastex}
\usepackage{emulateapj5}
\usepackage{epsfig}
 
\begin{document}

\title{
Structure in the Epsilon Eridani dusty disk caused by 
mean motion resonances with a 0.3 eccentricity planet
at periastron
}

\author{ A.~C. Quillen
\& Stephen Thorndike
}
\affil{Department of Physics and Astronomy, University of Rochester, Rochester, NY 14627; 
aquillen@pas.rochester.edu, slt3@alfred.edu
}

\begin{abstract}             

The morphology of the $\epsilon$ Eridani dust ring is 
reproduced by a numerical simulation of 
dust particles captured into the 5:3 and 3:2 
exterior mean-motion resonances with a 0.3 eccentricity $10^{-4}$ solar
mass planet
at periastron at a semi-major axis of 40 AU.
The morphology will differ when the planet is at apastron,
in about 140 years.  Moderate eccentricity planets
in outer extra-solar systems will cause observable variations in the
morphology of associated dusty rings.
\end{abstract}             


\section{Introduction}   

The Infrared Astronomical Satellite (IRAS) measured far infrared fluxes
from the vicinity of main sequence stars \citep{aumann}, including
the nearby systems Vega and Formalhaut which are of moderate
age, $10^8$ years, at the end of the era when rocky planets are expected
to form.  Recently,
submillimeter wavelength images of stars with
infrared excesses have revealed structure in some of these disks. 
Formalhaut has a dust ring with a central cavity \citep{holland}, Vega, $\beta$ Pictorus
and HR 4796A have secondary dust emission peaks and central cavities  
\citep{wilner,holland, jaya,koerner}, 
HR 4796A has a ring evident in scattered light from the central star
\citep{schneider},
and $\epsilon$ Eridani has a
60 AU dusty ring with peculiar 4 peaks of emission \citep{greaves}.
HR 4796A and $\beta$ Pictorus are younger, at ages of only $\sim 10^7$ years,
whereas $\epsilon$ Eridani (HR 1084) is a nearby 0.5-1 Gyr old K2V 
($0.8M_\odot$) star \citep{soderblom,song}.

Because of the short lifetime of the dust particles in these disks, 
and the scarcity of gas, the dust must continually
be replenished from collisions of larger orbiting bodies, possibly
analogous to Kuiper belt objects.  Hence the disks in the older systems
are commonly called debris disks. 
Dust particles in our solar system spiral toward the sun due to radiation
drag forces, known as Poynting-Robertson (P-R) drag and interactions with the solar wind
\citep{burns}.  Gravitational perturbations from planets 
can scatter the particles or capture them into orbital
resonances with planets (e.g., \citealt{dermott}).  
Since the lifetime of captured
particles can greatly exceed the lifetime of particles which
are not captured, resonant trapping can sculpt the dusty disk.

Structure predicted in the dust distribution in the Kuiper Belt
is primarily a result of resonance trapping \citep{liou99,amaya}.
\citet{wilner} suggested
that the two peaks seen at 1.3mm in the Vega disk could be explained
by the capture of dust particles
into the 2:1 exterior mean-motion resonance with a high eccentricity three Jupiter
mass giant planet at a semi-major axis of 40AU.
\citet{ozernoy} proposed that the morphology of the submillimeter
images of the $\epsilon$ Eridani system could be caused by 
dust particles trapped in high libration orbits associated
with the 3:2 mean motion resonances
with a $\sim 0.2$ Jupiter mass planet in a circular orbit.
While \citet{ozernoy} considered the morphology caused
by dust particles carefully placed in individual resonances, 
here we initially distribute
dust particles over a range of semi-major axes
and eccentricities and search for stable dust configurations
which can account for the four peaks 
observed in the submillimeter images of the $\epsilon$ Eridani system.

\section{Numerical Integration}

We numerically integrate the orbits of dust particles
using a conventional Burlisch-Stoer numerical scheme.
The forces on the particles are gravity from
a central star and one planet and P-R drag from the central star.   
The motion of the dust particles was restricted to the plane 
containing the planet and star.
The ratio of the force from radiation to that from the gravity
of the central star
is parametrized with the dimensionless parameter, 
$\beta$, which depends primarily on particle size \citep{burns}.
Length scales are given 
in units of the perturbing planet semi-major axis, $a_p$, and timescales
in units such that $GM_* =1$ where $M_*$ is the mass of the central star.
The ratio of the planet mass to the stellar mass is denoted as $\mu$.

Dust particles see a reduced radial force from the central
star because radiation pressure opposes gravity.  
This causes the location of the mean motion resonances with the planet 
to shift radially by a factor of $(1-\beta)^{1/3}$.  
However the magnitude of the P-R drag force 
caused by relativistic effects depends on the ratio of the particle
velocity to the speed of light, $c$, times $\beta$.
Because of our choice of units,
the speed of light, $c$, is in units of $\sqrt{GM_*/a_p}$.
We have adopted $c/\sqrt{GM_*/a_p}=5 \times 10^4$ which corresponds to having a stellar
mass of $1 M_\odot$ and a planet radius of 25AU.
Because of the weak dependence of $c$ on stellar mass and
planetary semi-major axis, our simulations should scale to outer solar
systems, such as the $\epsilon$ Eridani disk,
with the caveat that the particle drift rates should be 
a factor of a few times lower.

With the exception of \citet{wilner},
most simulations of dust dynamics in planetary systems have been
carried out for low eccentricity systems. 
As \citet{wilner} found,
high eccentricity planets can cause strong azimuthal density contrasts
in the distributions of objects trapped in outer resonances such
as the 2:1 and 3:1 mean motion resonances.
Typically we find that the azimuthal contrasts are lower in the 
outer resonances than those closer to the planet. 
Previous simulations have predicted that resonances close to
the planet such as
the  4:3, 3:2, 5:3, and 7:4 resonances efficiently
trap and hold dust particles in Kuiper belt \citep{liou99,amaya}.
We restrict our simulations to the region containing these resonances.

We began the simulations with 300 dust particles distributed randomly
in semi-major axis between 1.1 and 1.5 times the planet's semi-major axis.
Longitudes of periastron and mean longitudes were chosen randomly.
Dust particle eccentricities were chosen to lie between zero and 0.4.
We ran each simulation  for $2 \times 10^4$ planetary orbit periods.
The positions of all the particles were recorded 
every planetary year, consequently the phase of the planet was the same
during each output of the integration. 
We binned the particle positions counting the positions the particles 
at each output of the simulation.
The resulting spatial distributions and semi-major axis distributions
are shown in Figure 1 for the planet at periastron and at apastron.
All dust particles contributed to the time integrated dust distributions,
however,
particles which were quickly ejected from the system, or quickly impacted
the star or planet, contributed less to the integrated
dust distributions.     Particles which remained trapped in
resonances for long periods of time contributed substantially 
to the dust distribution.

In Figure 1d we show the particle distribution restricted
to semi-major axis between 1.2-1.3,  containing the 3:2 resonance,
and between 1.3-1.37, containing the 5:3 resonance, for the simulations shown 
in Figure 1a,b when the planet is at periastron and at apastron.
Only one family of resonant orbits dominates for both resonances
and the phase of the density peaks with respect to the planet 
depends on the planet's orbital phase.
The distribution of particles in the 3:2 resonance exhibits a three-fold
symmetry and that of the 5:3 exhibits a five-fold symmetry. 
When the planet is at periastron,
the sum of the particle distributions in the 3:2 and 5:3 resonances
exhibits four peaks similar to those observed in the $\epsilon$ Eridani system.

To simulate the effect of the decreasing radiation field
the dust distribution was weighted by $r^{-2}$ where $r$ is
its radial distance from the star.
The resulting predicted intensity image for a planet at periastron,
and smoothed to the approximate resolution of the submillimeter observations
is shown in Figure 2 with the submillimeter image by \citet{greaves} for comparison.
The dusty disk is tilted assuming a disk inclination of $40^\circ$ 
(where $0^\circ$ corresponds to face on).  This is somewhat
higher than the $25^\circ$ estimated by \citet{greaves} from the ring
but consistent with the $30^\circ \pm 15^\circ$ estimated from optical
line data \cite{saar}.

The morphology of the dust simulation shown in Figure 2 exhibits two dominant peaks
at mean longitudes of approximately $\pm 3 \pi/5$ from the planet which correspond
fairly well to the two dominant peaks south east and west of the star 
observed in the submillimeter map.  Near the planet, a weaker arc 
is displayed in the simulation, which corresponds to the weaker arc of emission
observed to the north of the star in the submillimeter map.
However, the dust emission predicted opposite to the planet
is more distant from the star than that observed
south of the star in the submillimeter image.

In Figure 1d we show particle distributions for restricted ranges of
semi-major axes.  When the planet is at periastron, density enhancements
caused by  the 3:2 resonance add to those from the 5:3 causing
the two dominant peaks seen in Figure 1a.  When the planet is at apastron,
the concentrations caused by the 3:2 resonance are out of phase with
those in the 5:3.  Some substructure is also caused by the 7:4 resonance
in Figures 1a,b.

In Figures 1 and 2 we have given axes in units of the planets semi-major 
axis.  The radius of the ring was estimated to be approximately 60 AU which would
correspond to a planet semi-major axis
of about 2/3 this or 40 AU.  Assuming a stellar mass of 0.8 $M_\odot$,
the planet orbital period is 280 years.  The total time of our simulations
($2 \times 10^4$ times the planet period) corresponds to $6 \times 10^7$ years.

Dust particles can be trapped in a $k+q:k$ exterior principle mean-motion resonance 
with resonant arguments of the form
\begin{equation}
\phi = (k+q)\lambda - k \lambda_p - p\varpi + (p-q) \varpi_p
\end{equation}
where $p$ is an integer between 0 and $q$,
$\lambda$ and $\lambda_p$ are the mean longitudes of the particle
and planet respectively,  and
$\varpi$ and $\varpi_p$ are the longitudes of periastron.
We consider the particle to be trapped in the resonance when $\phi$
librates about a particular fixed value.
Take the planet's longitude of periastron as a reference angle
($\varpi_p =0$).
In resonance the resonance angle $\phi \approx n\pi $ where $n$ is odd
if $q$ is even and $n$ is even otherwise.

For the mean longitude of a dust particle very close to the planet to be strongly locked
to that of the planet, the resonance responsible is likely to be the $p=0$
one.  In this case
\begin{equation}
\lambda = { n \pi + k \lambda_p \over k + q}
\end{equation}
For the 3:2 resonance and when the planet is at periastron ($\lambda_p= \omega_p=0$),
we find dust particles near $\lambda = 0, \pm 2\pi/3$.  When the 
planet is at apastron ($\lambda_p=\pi$), $\lambda$ is the same.
This only looks out of phase in Figure 1d because we have rotated the images
so that the longitude of the planet is fixed.
When the resonant argument $\phi$ is fixed, we will only
find particles with these longitudes, causing the concentrations
we see in Figure 1.
For the 5:3 resonance, when the planet is at periastron,
$\lambda = \pi, \pm 3\pi/5, \pm \pi/5$, and when the planet is
at apastron,  $\lambda = 0, \pm 2 \pi/5, \pm 4\pi/5$.   This five-fold symmetry
is also consistent with the concentrations seen in Figure 1d.

In our simulations
we also found particles populating resonances with other values of $p$,
however these tended to have shorter lifetimes and so contributed less
to the final particle distributions.
When $p\neq 0$, higher particle densities arise at angles when the particles 
reach apastron or periastron (see discussion in \citealt{wilner}).
For example, $p=1$ is favored for the 3:2 resonance
when the planet is at low eccentricity causing two peaks in the particle
distribution rather than three, which is what is seen in the simulations
of \citep{liou99}.

\section{Summary and Discussion}

We have found that the morphology of dusty material 
trapped in exterior resonances with a planet can be strongly
dependent upon the eccentricity and orbital phase of the planet.
The morphology of the $\epsilon$ Eridani dust ring seems to
be reproduced by dust particles captured into the 5:3 and 3:2 
exterior resonances with a moderate eccentricity, $e_p \sim 0.3$, planet
near periastron.  
When the planet was more massive ($\mu > 5 \times 10^{-4}$) 
or more highly eccentric (greater than 0.4),
the resonances closest to the planet did not capture and hold
particles as efficiently.  When the planet mass was below a Saturn 
mass, the 4:3 resonance also contributed to the particle distribution resulting
in dominant symmetric four peaks.  The asymmetries
observed in the $\epsilon$ Eridani image were not observed in 
the dust distribution so we consider planets
less massive than Saturn unlikely to explain
the particle distribution.  When the planet eccentricity was below 0.15,
the azimuthal density variations in the dust distribution were too low to 
account for the morphology in the $\epsilon$ Eridani disk.

Our model differs from that proposed by \citet{ozernoy} in a number
of ways.  The period of the orbital planet in our model is 
280 years which should cause the pattern to revolve
by about the star by $\sim 1.3^\circ {\rm yr}^{-1}$, faster
than the $\sim 0.7^\circ {\rm yr}^{-1}$ estimated by \citet{ozernoy},
where the model planet semi-major axis is $\sim 60$AU rather than at $\sim 40$AU.
Our model planet is located to the north of the star, rather
than to the west of the star.  Our model planet mass
is similar to theirs but at a moderate eccentricity.  
Our model dust concentrations are a result of segregation in the phase
of a resonant angle, rather than caused by a large libration
amplitude.  Furthermore, because
of the eccentricity of the planet, our model predicts
that the morphology of the dusty ring will vary, as well as revolve 
as the planet orbits about the star.

The initial conditions of our simulations cause many of the particles
to begin trapped in the resonances and at fairly low eccentricity.
It is possible that low eccentricity planetesimals exist in the system
and that they are the source of the dust particles
that we see.   Alternatively the planetesimals in the system 
are further out and the dust particles become trapped in the 
resonances as they spiral in toward the planet.
Further simulations would be required to differentiate between
these possibilities.
In either case, resonant capture into these resonances is less likely
and less prolonged 
when the planet mass or eccentricity is high.  
If the planet mass is too low then it cannot be responsible for clearing
a gap or central region in the dust distribution.  Additional and more
massive planets would be required to do this.

It is possible that 
high eccentricity planets are common in the outskirts of extra-solar
systems.  If so then the resulting dust distributions would
not only revolve \citep{ozernoy}, but will also
be dependent upon the orbital phase of the planets. 
This is an exciting prospect because there would be variations in 
the dust morphology on observable timescales.

\begin{figure*}
\vspace{16.0cm}
\includegraphics{h3_y.ps}
\includegraphics{h4_y.ps}
\includegraphics{h3_s.ps}
\includegraphics{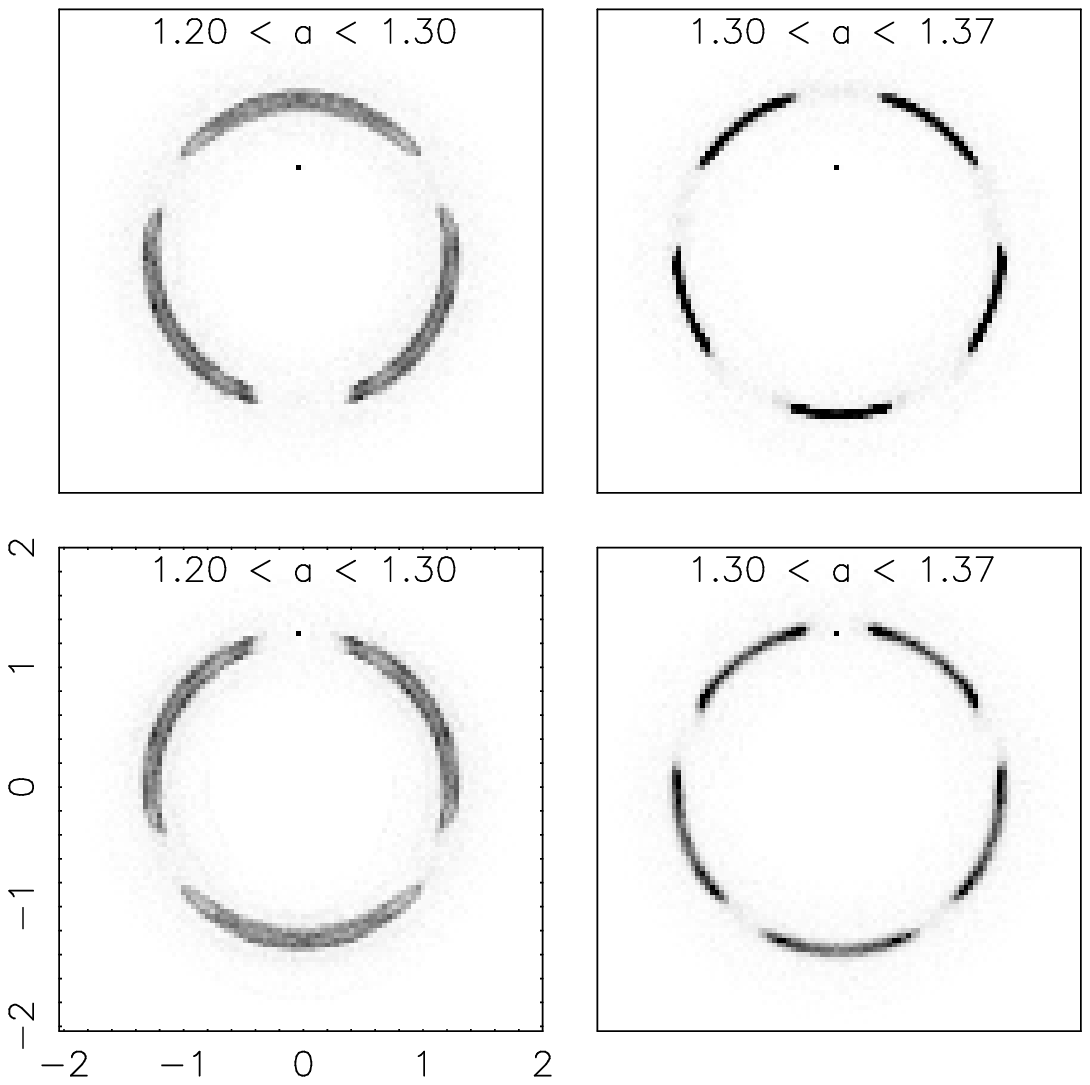}
\caption[]{
a) The dust distribution for a planet
at periastron with ratio of planet mass to stellar mass $\mu=10^{-4}$,
and eccentricity of $e_p=0.3$.
The dust particles have $\beta=0.1$.
The star and planet are denoted as black dots at the origin 
and at a radius of 0.7 from the origin, respectively.
Axes are given in units of the planet semi-major axis.
b) Same as a) but the planet is at apastron.
c) Semi-major axis distribution for the particle distribution shown in a.
d) Particle distributions are shown for limited ranges of semi-major axes
for the simulations displayed in Figures 1a,b.  Particles in the 3:2 resonance
are shown for the planet at periastron (top left)
and at apastron (bottom left).   Particles in the 5:3 resonance
are shown for the planet at periastron (top right) and at apastron
(bottom right).
}
\end{figure*}

\begin{figure*}
\vspace{10.0cm}
\caption[]{
On the right is shown 
the simulated intensity distribution  based on the simulation shown in Figure 1a,
for a 0.3 eccentricity planet at periastron of planet to stellar mass ratio $10^{-4}$.
For comparison we show on the left
the $850\mu$m submillimeter emission map of the $\epsilon$ Eridani system.
(this is Figure 1 by \citealt{greaves}).
The location of the star is denoted in both panels as a star.
For the simulated intensity distribution we have chosen
a disk inclination of $40^\circ$ where $0^\circ$ is face on.
The dust distribution has been weighted by $r^{-2}$ to simulate
the intensity field from the star and then smoothed to the approximate
resolution of the submillimeter images.
We estimate that the planet is currently located to the north of the star 
(denoted as a filled circle in the right hand panel)
at a declination of about $-9^d27^m20^s$ (epoch 1998 as from Figure 1 of \citealt{greaves}).
Axes for the simulation are given in units of the semi-major axis
of the planet.
}
\end{figure*}

\acknowledgments                             
This work would not have been carried out without helpful
discussions with Stephen Thorndike, Joel Green and Dan Watson.

%
%


  
{}


\begin{thebibliography}{}

\bibitem[Aumann et al.(1984)]{aumann}
Aumann, H.~H. et al.~1984, ApJ, 278, L23

\bibitem[Burns et al.(1979)]{burns}
Burns, J.~A., Lamy,  O.~L., Soter, S.~1979, Icarus, 40, 1


\bibitem[Dermott et al.(1994)]{dermott}
Dermott, S.~F., Jayaraman, S., Xu, Y.~L., Gustafson, B.~A.~S., \& Liou, J. C.
1994, Nature, 369, 719


\bibitem[Greaves et al.(1998)]{greaves}
Greaves, J. S., et al.~1998, ApJL, 506, L133


\bibitem[Holland et al.(1998)]{holland}
Holland, W.~S., et al.~1998, Nature, 392, 788 

\bibitem[Jayawardhana et al.(1998)]{jaya}
Jayawardhana, R., Fisher, S., Hartmann, L., Telesco, C.,
Pina, R., \& Fazio, G.~1998, ApJ, 503, L79

\bibitem[Koerner et al(1998)]{koerner}
Koerner, D.~W., Ressler, M.~E., Werner, M.~W., \& Backman, D.~E.
1998, ApJ, 503, L83

\bibitem[Murray \& Holman(1997)]{murray}
Murray, N., \& Holman, M.~1997, AJ, 114, 1246




\bibitem[Liou \& Zook(1999)]{liou99}
Liou, J.-C., \& Zook, H.~A.~1999, AJ, 118, 580

\bibitem[Moro-Martin \& Malhotra(2002)]{amaya}
Moro-Martin, A. \& Malhotra, R.~2002, AJ, in press, (astroph/0207350)

\bibitem[Ozernoy et al.(2000)]{ozernoy}
Ozernoy, L.~M., Gorkavyi, N.~N., Mather, J.~C.,\& Taidakova, T.~A.\
2000, ApJ, 537, L147

\bibitem[Saar \& Osten(1997)]{saar}
Saar, S.~H., \& Osten, R.~A.~1997, MNRAS, 284, 803

\bibitem[Schneider et al.(1999)]{schneider}
Schneider, G., et al.~1999, ApJ, 513, L127

\bibitem[Soderblom \&  Dappen(1989)]{soderblom}
Soderblom, D., \&  Dappen, W.~1989, ApJ, 342, 945

\bibitem[Song et al.(2000)]{song}
Song, I., Caillault, J.-P., Barrado y Navascu{\'e}s, D., 
Stauffer, J.~R., \& Randich, S.\, 
2000, ApJL, 533, L41

\bibitem[Wilner et al.(2002)]{wilner}
Wilner, D.J., Holman, M.J., Kuchner, M.~J., \& Ho, P.T.P.~2002, ApJ, 569, L115


\end{thebibliography}
\end{document}